\documentclass[twocolumn,showpacs,aps,prl,amsmath,amssymb,superscriptaddress]{revtex4}
\usepackage{graphicx}
\usepackage{dcolumn}
\usepackage{bm}
\usepackage{amssymb}
\usepackage[german, english]{babel}
\usepackage[usenames]{color}
\usepackage{epsfig}

\begin{document}

\title{Pairing-gap, pseudo-gap, and no-gap phases in the radio-frequency spectra of a trapped unitary $^{6}$Li gas}

\author{P. Pieri}
\affiliation{Dipartimento di Fisica, Universit\`{a} di Camerino, 62032 Camerino, Italy}

\author{A. Perali}
\affiliation{Dipartimento di Fisica, Universit\`{a} di Camerino, 62032 Camerino, Italy}

\author{G. C. Strinati}
\affiliation{Dipartimento di Fisica, Universit\`{a} di Camerino, 62032 Camerino, Italy}

\author{S. Riedl}
\altaffiliation[Present address: ]{Max-Planck-Institut f\"ur Quantenoptik, Garching, Germany.}
\affiliation{Institut f\"ur Experimentalphysik und Zentrum f\"ur Quantenphysik, Universit\"at Innsbruck, 6020 Innsbruck, Austria}
\affiliation{Institut f\"ur Quantenoptik und Quanteninformation, \"Osterreichische Akademie der Wissenschaften, 6020 Innsbruck, Austria}

\author{M. J. Wright}
\altaffiliation[Present address: ]{Harvard University, Cambridge, USA.}
\affiliation{Institut f\"ur Experimentalphysik und Zentrum f\"ur
Quantenphysik, Universit\"at Innsbruck, 6020 Innsbruck, Austria}

\author{A. Altmeyer}
\altaffiliation[Present address: ]{Deutsche Bahn Mobility Logistics AG, Berlin, Germany.}
\affiliation{Institut f\"ur Experimentalphysik und Zentrum f\"ur
Quantenphysik, Universit\"at Innsbruck, 6020 Innsbruck, Austria}
\affiliation{Institut f\"ur Quantenoptik und Quanteninformation,
\"Osterreichische Akademie der Wissenschaften, 6020 Innsbruck, Austria}

\author{C. Kohstall}
\affiliation{Institut f\"ur Experimentalphysik und Zentrum f\"ur
Quantenphysik, Universit\"at Innsbruck, 6020 Innsbruck, Austria}

\author{E. R. S\'anchez Guajardo}
\affiliation{Institut f\"ur Experimentalphysik und Zentrum f\"ur
Quantenphysik, Universit\"at Innsbruck, 6020 Innsbruck, Austria}

\author{J. {Hecker Denschlag}}
\altaffiliation[Present address: ]{Universit\"at Ulm, Germany.}
\affiliation{Institut f\"ur Experimentalphysik und Zentrum f\"ur
Quantenphysik, Universit\"at Innsbruck, 6020 Innsbruck, Austria}

\author{R. Grimm}
\affiliation{Institut f\"ur Experimentalphysik und Zentrum f\"ur
Quantenphysik, Universit\"at Innsbruck, 6020 Innsbruck, Austria}
\affiliation{Institut f\"ur Quantenoptik und Quanteninformation,
\"Osterreichische Akademie der Wissenschaften, 6020 Innsbruck, Austria}

%\date{(\today)}


\begin{abstract}
Radio frequency spectra of a trapped unitary $^{6}$Li gas are reported and analyzed in terms of a theoretical approach that includes both final-state and trap effects.
The different strength of the final-state interaction across the trap is crucial for evidencing two main peaks associated with two distinct phases residing in different trap regions.
These are the pairing-gap and pseudo-gap phases below the critical temperature $T_{c}$, which evolve into the pseudo-gap and no-gap phases above $T_{c}$.
In this way, a long standing puzzle about the interpretation of rf spectra for $^{6}$Li in a trap is solved.
\end{abstract}

\maketitle

Radio frequency (rf) spectroscopy has proved a valuable tool for exploring physical properties of fermionic ultra-cold atoms
\cite{Jin-2003,Ketterle-2003,Grimm-2004,Ketterle-2007,Schunck-2007,Ketterle-2008,Ketterle-2008b,Jin-2008}.
Interpretation of rf spectra as to extract properties of the single-particle excitations, however, is quite involved, in particular as it requires one to disentangle 
effects of the trapping potential and of interactions in the initial and final states of the rf transition. 

After rf spectra were first proposed in Ref.\cite{Grimm-2004} to signal the presence of a pairing gap for ultra-cold $^{6}$Li Fermi atoms at low enough temperature, the experimental emphasis 
has focused in eliminating trap effects via tomographic techniques that select the contributions from specific trap portions \cite{Ketterle-2007}, and in reducing the final-state interaction
between the atom excited by the rf pulse and the remaining atoms \cite{Ketterle-2008}.
Spectra for $^{40}$K, which are not appreciably affected by final-state effects, have been recorded for the whole trap in a momentum-resolved way \cite{Jin-2008}, mimicking the angular-resolved 
photoemission spectroscopy widely used in the context of high-Tc superconductivity.
From the theoretical side, the spectra of Refs.\cite{Ketterle-2007,Ketterle-2008} were interpreted by accounting for the role played by final-state effects in a homogeneous gas \cite{PPS-2008,PPS-2009}, while 
those of Ref.\cite{Jin-2008} require one to consider trap but no final-state effects.
In all these cases the rf spectra show a single feature \cite{bound-state} with an asymmetric shape.

Yet, it remains intriguing that the presence of two peaks in the original rf spectra of Ref.\cite{Grimm-2004} at temperatures about 
$T_{c}$ could not be accounted for in a fully satisfactory way thus far. 
In Refs.\cite{Torma-2004,Levin-2005} the two peaks were identified as coming from paired and non-paired atoms in different trap regions, but final-state effects were not included in these works. When applied to a homogeneous system,  the same theoretical approach predicts two peaks in the rf spectra over an extended temperature range \cite{Levin-2009}, in contrast with the tomographic data of Ref.\cite{Ketterle-2008b} 
(see in particular the supplemental material therein).
In addition, the work of Ref.\cite{MBS-2008} yields a single peak for a homogeneous system, in agreement with these tomographic data, but gives rise to two peaks in a trapped Fermi gas only for imbalanced spin populations.
We shall show that the puzzle posed by the original rf spectra of Ref.\cite{Grimm-2004} can be solved when final-state and trap effects
are taken into account simultaneously.

In this Letter, we present rf spectra of a unitary trapped $^{6}$Li gas with improved control of the experimental parameters with respect to Ref.\cite{Grimm-2004}. At the same time, we provide a detailed 
interpretation of these spectra on the basis of the theoretical approach of Refs.\cite{PPS-2008,PPS-2009}, which is here extended by a local-density approximation (LDA) to account for the trap and 
by proper inclusion of final-state effects
also for non-condensed pairs in the superfluid phase. 

The main results of our analysis are as follows:

\noindent
(i) We are able to confirm the interpretation given at an intuitive level already in Ref.\cite{Grimm-2004} for the two peaks in the rf spectra above (but quite close to) $T_{c}$, as associated with free atoms in the outer part of the trap and with interacting atoms in the pseudo-gap regime in the inner part of the trap. 
[In the present context, a pseudo-gap corresponds to a suppression of the spectral weight about zero energy detuning.]
We have verified that only one peak results instead when final-state effects are not taken into account;

\noindent
(ii) We extend the analysis below $T_{c}$ and identify the two main structures present in the rf spectra as being associated with condensed Cooper pairs (a pairing-gap phase) and with non-condensed Cooper pairs (a pseudo-gap phase).
In this case, a minor structure in the calculated rf spectra can be associated with free atoms residing in the outer part of the trap (a no-gap phase);

\noindent
(iii) We verify through a shell analysis, both above and below $T_{c}$, that the different peaks present in the rf spectra originate from different trap regions.

The experimental starting point of our new rf spectra is a deeply degenerate, strongly interacting Fermi gas of $N = 4 \times 10^5$ $^{6}$Li atoms in an optical trap. The sample is prepared in a balanced spin mixture of the two lowest hyperfine sub-levels $|1\rangle$ and $|2\rangle$.
The basic preparation procedures are described in our previous work \cite{Jochim-2003,Grimm-2004,Wright-2007}. 
The mean trap frequency is $\bar{\omega} = 2\pi \times 196$\,Hz \cite{scissors}, leading to a Fermi temperature $T_F = (\hbar \bar{\omega}/k_B)\,(3N)^{1/3} = 1.0\,\mu$K and a corresponding Fermi wavenumber $k_F = (3800 a_0)^{-1}$, where $\hbar$ is the reduced Planck constant, $k_B$ the Boltzmann constant, and $a_0$ the Bohr radius.
We set the magnetic field at the center of the broad 834-G Fano-Feshbach resonance.
Our thermometry is based on measuring an effective temperature parameter $\tilde{T}$ and an empirical conversion into the real temperature $T$ 
\cite{Wright-2007,Kinast-2005,Kinast-PhD,Stajic-2005}.

To record the rf spectra we follow the procedure described in Ref.\cite{Grimm-2004}. By a weak 0.5\,s pulse, we drive transitions from $|2\rangle$ to the empty state $|3\rangle$ and monitor the loss of atoms in $|2\rangle$. After the rf pulse no population can be detected in $|3\rangle$, which results from fast three-body decay in a three-component spin mixture \cite{Ottenstein-2008,Huckans-2009}. 
With the scattering lengths $a_i \rightarrow \pm \infty$ and $a_f = -3310\,a_0$ \cite{Bartenstein-2005} for the initial channel $|1\rangle$-$|2\rangle$ and the final channel $|1\rangle$-$|3\rangle$, the corresponding couplings are $(k_{F} a_{i})^{-1}=0$ and $(k_{F} a_{f})^{-1}=-1.15$.

% Figure 1
\begin{figure}[t]
\includegraphics[angle=0,width=8.5cm]{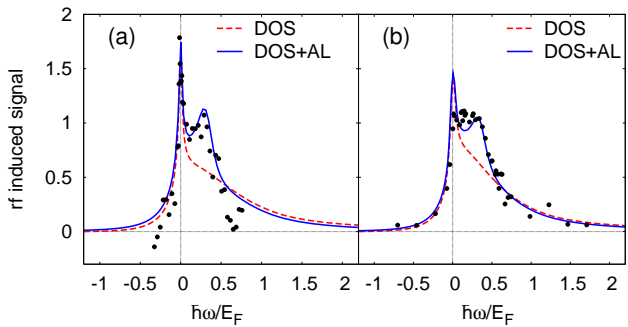}
\caption{Experimental rf spectra of a trapped $^{6}$Li gas near $T_{c}$ (circles) are compared with
theoretical calculations (DOS+AL) or (DOS) which include or neglect final-state effects.
(a) Data near unitarity (822G) reproduced from Ref.\cite{Grimm-2004}; 
(b) New data at unitarity (834G).}
\label{fig1}
\end{figure}

The main differences of the present measurements to the early rf spectra in Ref.\cite{Grimm-2004} are a greatly improved control of all experimental parameters, better knowledge of the 
resonance position \cite{Bartenstein-2005}, and a better thermometry.

We begin our analysis by reconsidering the data shown in the central panel of Fig.1 of Ref.\cite{Grimm-2004}, corresponding to a magnetic field of 822G near resonance.
We report those data in Fig.\ref{fig1}(a), where the energy detuning $\hbar\omega$ with respect to the atomic transition is given in units of the Fermi energy $E_{F} = k_{B} T_{F}$.
The original estimate $T_{F} = 3.4\,\mu$K \cite{Grimm-2004} has been subject to revision, yielding $T_{F} = 2.7\,\mu$K,
because of an improved characterization of the original trap parameters and particle number. 
The temperature at which the data were taken is estimated to be $0.3 T_{F}$, which is close to $T_{c}$ \cite{temperature},
while $(k_{F} a_{i})^{-1}=0.07$ and $(k_{F} a_{f})^{-1}=-0.7$.
This accounts for the differences between Fig.\ref{fig1}(a) and Fig.\ref{fig1}(b) which reports the new data.

The double-peak structure in the rf spectrum of Fig.\ref{fig1}(a) was already regarded in Ref.\cite{Grimm-2004} as an indication of the ``coexistence of unpaired and paired atoms''.
Here, we provide the desired quantitative explanation for the data of Fig.\ref{fig1}(a), by calculating the rf spectrum above $T_{c}$ extending the theory of Ref.\cite{PPS-2009} to account for the contributions 
from different shells in the trap through LDA.
Calculations are presented both in the absence (broken line) and presence (full line) of final-state effects.
In both cases, the theory contains no adjustable parameters.
The first calculation corresponds to a density-of-state (DOS) approximation for excitations in the continuum which includes pairing in the initial state via a self-energy contribution, and the second one to the further inclusion of final-state effects 
by Aslamazov-Larkin (AL) type processes \cite{PPS-2009} where the interaction between the excited atom and the mate left behind is taken into account. 

In this and the following figures, the vertical scale is set by normalizing the theoretical  spectra to a unit area (sum rule) \cite{PPS-2008}, while a corresponding normalization cannot be exploited for the experimental spectra owing to uncertainties in the large-$\omega$ tails.
 We then make the height of the right experimental peak to coincide with the theoretical prediction.

The DOS calculation (broken line), which neglects final-state effects, cannot reproduce the broad right peak present in the experimental data, which is of most interest being associated with nontrivial pairing effects.
Within the DOS approximation, pairing effects lead only to an asymmetric peak near $\omega = 0$ with a slow decay
$\propto \omega^{-3/2}$ for large $\omega$. 
It is only when final-state effects are included in the DOS+AL calculation that the presence of the second peak at a finite frequency can be accounted for, both for its position and width.

% Figure 2
\begin{figure}[t]
\includegraphics[angle=0,width=8.5cm]{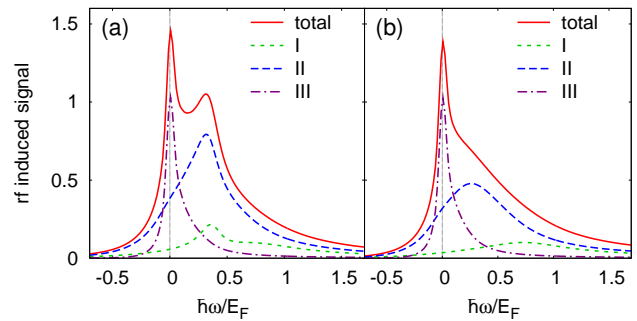}
\caption{Shell analysis of rf spectra for a trapped unitary Fermi gas at $T_c$, (a) with and (b) without final-state effects 
when $(k_{F} a_{f})^{-1}=-1.15$. The regions I, II, and III are defined in the text.}
\label{fig2}
\end{figure}

It was shown in Refs.\cite{PPS-2008,PPS-2009} that for a homogeneous system final-state effects pull the oscillator strength of the rf transition toward threshold ($\omega=0$) and make the tail to decay faster $\propto \omega^{-5/2}$ (the total area of the spectrum remaining constant).
In a trap, final-state effects are stronger where the density is larger, due to the larger values of the local final-state coupling 
$(k_{F}(r) a_{f})^{-1}$ and the smaller values of the local reduced temperature $T/T_F(r)$.
Final-state effects therefore pull the oscillator strength toward threshold, in a more marked way in the trap center where the pseudo-gap is larger, and in a less marked way when approaching the trap edge where the pseudo-gap is progressively reduced.
The net effect is that the inner and intermediate regions of the trap ($\mathrm{I}$ and $\mathrm{II}$ in the shell analysis of Fig.\ref{fig2}(a)) produce partial peaks at essentially the same frequency in the rf spectra.
On the other hand, in the absence of final-state effects there is no pulling toward threshold which can compensate for the spread of pseudo-gap values associated with different trap shells, with the net effect of producing only a broad shoulder in the total rf spectra
(cf. Fig.\ref{fig2}(b)).
In both cases, the peak at zero detuning results from free atoms which reside in the outer part of the trap (region $\mathrm{III}$
in Figs.\ref{fig2}(a) and \ref{fig2}(b)).
In the shell analysis we have taken regions I, II, and III to correspond to $r/R_{F}<0.4$, $0.4< r/R_{F}< 0.8$, and $r/R_{F} > 0.8$, in the order, where $R_{F}=[2E_{F}/(m\bar{\omega}^{2})]^{1/2}$ is the Thomas-Fermi radius \cite{shell-structure}.

This interpretation is supported by new data exactly at resonance near $T_{c}$, which are 
reported in Fig.\ref{fig1}(b) together with the theoretical calculations in the normal phase, with and without final-state effects.
Note that the two theoretical peaks are less pronounced in Fig.\ref{fig1}(b) than in Fig.\ref{fig1}(a).
This is because the final coupling $(k_{F} a_{f})^{-1}=-1.15$ of Fig.\ref{fig1}(b) is consistently weaker than $(k_{F} a_{f})^{-1}=-0.7$ of Fig.\ref{fig1}(a).

Comparison between theory and experiment is again rather remarkable, the only discrepancy appearing in the height of the free-atom peak which is somewhat decreased in the experimental data \cite{footnote-loss}. 
Comparison at higher temperatures yields further very good agreement between experiment and theory \cite{supplemental-material}.

\begin{figure}[t]
\includegraphics[angle=0,width=8.5cm]{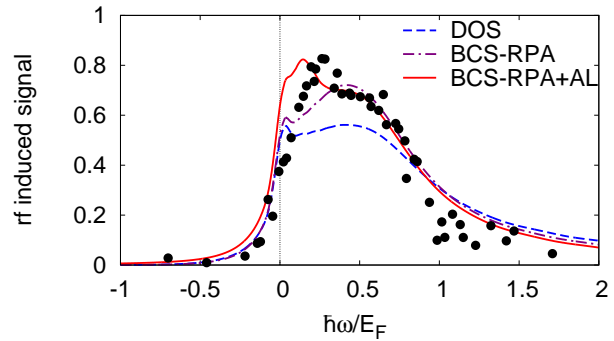}
\caption{Experimental rf spectra of a trapped $^{6}$Li gas below $T_{c}$ (circles) are compared with
theoretical calculations with (BCS-RPA, AL) or without (DOS) final-state effects.}
\label{fig3}
\end{figure}

We also obtained new experimental data for the superfluid phase below $T_{c}$, as reported in Fig.\ref{fig3} for $T=0.1T_{F}$. 
Three distinct calculations are reported in this case: (a) The DOS approximation extended to the superfluid phase (broken line); (b) The corresponding inclusion of final-state effects for the \emph{condensed} pairs through the BCS-RPA approximation of Ref.\cite{PPS-2008} (broken-dotted line); (c) The further inclusion of final-state effects for the \emph{non-condensed} pairs through an extension of the AL contribution of Ref.\cite{PPS-2008} to temperatures below $T_{c}$ (full line).
At this temperature, the order parameter is non-vanishing for $r<r_{0}=0.63R_{F}$.
The DOS approximation is used for both $r < r_{0}$ and $r > r_{0}$, the BCS-RPA approximation is included only for $r < r_{0}$ to account for final-state effects associated with \emph{condensed} Cooper pairs, while the AL contribution that accounts for final-state effects associated with \emph{non-condensed} Cooper pairs is included both in the normal phase for $r > r_{0}$ and superfluid phase for $r < r_{0}$.

% Figure 3

The agreement between the experimental data below $T_{c}$ and the most complete theoretical calculation (full line) appears quite good, in particular for the overall shape, width, and wings. 
The theory predicts the presence of a peak at about $0.2 E_{F}$ and of a shoulder at about $0.5 E_{F}$.
This structure is also clearly visible in the experimental data, with a slight discrepancy in the position of the main peak which is shifted to about $0.3 E_{F}$.
This discrepancy could again result from a partial loss of the free-atom contribution in the experimental data \cite{footnote-loss}.
As a consequence, the theoretical spectrum has more weight than the experimental one near zero energy.
Note the slow decay of the tail of the DOS calculation \cite{supplemental-material}.

\begin{figure}[t]
\includegraphics[angle=0,width=6.5cm]{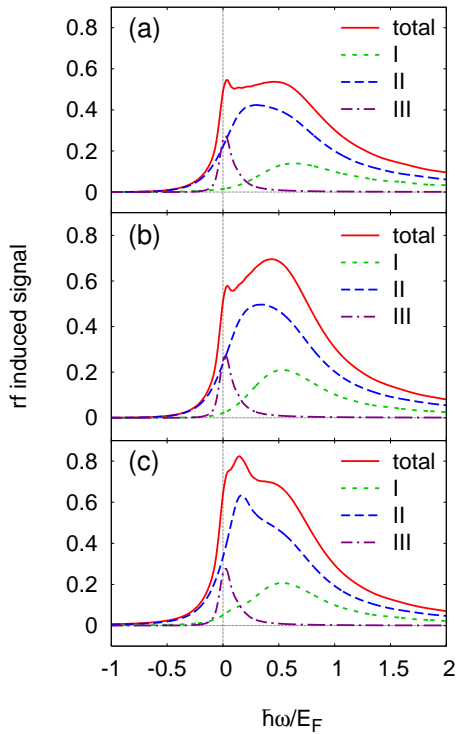}
\caption{Shell analysis of rf spectra of a unitary trapped Fermi gas below $T_c$ (for $(k_{F} a_{f})^{-1}=-1.15$):
(a) Without final-state effects; With final-state effects (b) only for condensed pairs and (c) also for non-condensed pairs. 
As in Fig.\ref{fig2} the curves refer to different regions in the trap.}
\label{fig4}
\end{figure}

A shell analysis again helps in understanding the origin of the features in Fig.\ref{fig3}.
Accordingly, we report in Fig.\ref{fig4} the shell analysis of the rf spectra below $T_{c}$, using the same three regions 
of Fig.\ref{fig2}.
The three panels of Fig.\ref{fig4} correspond to the three theoretical approximations specified in the caption of Fig.\ref{fig3}.
This analysis shows that:
(i) The contribution from the inner part of the trap (region $\mathrm{I}$) is essentially the same in panels (b) and (c), thus showing that only a small number of non-condensed pairs can be present there;
(ii) In the intermediate region $\mathrm{II}$ of the trap final-state effects are important both for condensed and non-condensed Cooper pairs;
(iii) The contribution from the outer part of the trap (region $\mathrm{III}$) remains the same within the three approximations, because in this region final-state effects are irrelevant.

This shell analysis shows that the two main structures present in the total spectrum obtained by the most complete calculation (full line in Fig.\ref{fig4}(c)) correspond to non-condensed Cooper pairs residing in the intermediate trap region $\mathrm{II}$ (left peak)
and to condensed Cooper pairs residing in the inner trap region $\mathrm{I}$ (right shoulder).
A third structure near zero energy is also noticed in the full spectrum of Fig.\ref{fig4}(c), which results from free atoms in 
region $\mathrm{III}$ of the trap corresponding to a no-gap phase.

% Figure 4

With the support of the shell analyses of Fig.\ref{fig2} (above $T_{c}$) and of Fig.\ref{fig4} (below $T_{c}$), we conclude that
the trap brings about a favorable circumstance: Distinct phases are present at the same time in different regions of the trap, 
and final-state effects render them observable as distinct features in the rf spectra.

In conclusion, we have presented an extended comparison between a new set of experimental rf spectra of a trapped unitary 
$^{6}$Li gas, both above and below $T_{c}$, with improved control of the experimental parameters, and the corresponding theoretical calculations that include final-state on top of trap effects.
These two effects conspire constructively with each other, and provide the solution to a long standing puzzle about the interpretation of rf spectra for $^{6}$Li in a trap.
They evidence the simultaneous presence of distinct phases, above as well as below $T_{c}$:
The pseudo-gap and no-gap phases above $T_{c}$; the pairing-gap, pseudo-gap, and no-gap phases below $T_{c}$. 

%%%%%%%%%%%%%%%%%%%%%%%%%%%%%%%%%

This work was partially supported by the Italian MIUR under Contract PRIN-2007 ``Ultracold Atoms and Novel Quantum Phases'', and by the Austrian Science Fund (FWF) within SFB 15 and SFB 40. 
M.J.W.\ was supported by a Marie Curie Incoming International Fellowship within the 6th EC Framework Program.
                                                                                                                                                                                                                                                                                                                                                                                                           
%%%%%%%%%%%%%%%%%%%%%%%%%%%%%%%%%
% Bibliography

%\newpage
\vspace{0.5cm}

\begin{center}
{\bf Supplemental material}
\end{center}
% to: ``Pairing-gap, pseudo-gap, and no-gap phases in rf spectra of a trapped unitary $^{6}$Li gas''}

\begin{center}
{\bf Comparison at high temperatures}
\end{center}

\begin{figure}[h]
\includegraphics[angle=0,width=6cm]{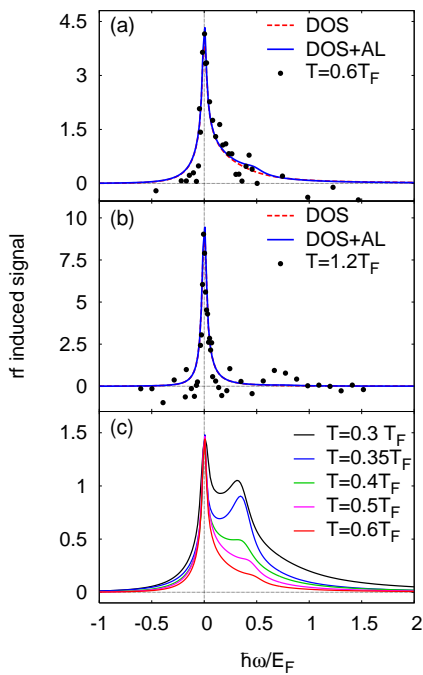}
\caption{Temperature evolution of the rf spectra of a trapped unitary $^{6}$Li gas.
Experimental spectra (circles) for (a) $T=0.6T_{F}$ and (b) $T=1.2T_{F}$ are compared with theoretical calculations 
which include (DOS+AL) or neglect (DOS) final-state effects. In (c) the temperature evolution of the theoretical spectra evidences 
the progressive disappearance of the two-peak structure for increasing temperature (here, the spectra were rescaled to share the height of the left peak).}
\label{figS1}
\end{figure}

\noindent
To test the evolution of the pseudo-gap phase and its fate with increasing temperature, we report in 
Figs.\ref{figS1}(a) and \ref{figS1}(b) the rf spectra taken at temperatures $0.6T_{F}$ and $1.2T_{F}$, respectively, with the corresponding DOS and DOS+AL calculations appropriate to the normal phase.

% Figure S1

The peak at larger frequencies associated with the pseudo-gap phase is seen to disappear for increasing temperature, while the remaining peak at zero detuning looses its pronounced asymmetry toward high frequencies.
To evidence this effect more clearly, we report in Fig.\ref{figS1}(c) a theoretical DOS+AL calculation for five temperatures in the normal phase between $0.3T_{F}$ and $0.6T_{F}$, where the progressive disappearance of the peak at larger frequencies for increasing temperature results evident.

\begin{center}
{\bf The slow frequency tail of the DOS calculation}
\end{center}

\noindent
A comment is in order about the comparison which results in Fig.3 of the main text, between the experimental data  and the DOS calculation (that neglects final-state effects).
At a first sight it may, in fact, look like that the DOS calculation by itself could account for the overall features that are present in the experimental data.

% Figure S2
\begin{figure}[h]
\includegraphics[angle=0,width=8.0cm]{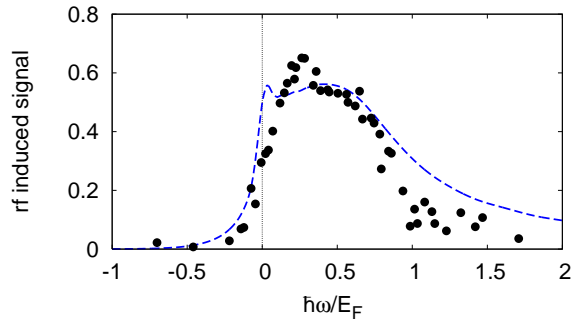}
\caption{The dashed line represents the DOS calculation of Fig.3 of the main text, while the circles are the corresponding experimental data that have been rescaled vertically as to match the theoretical curve near $0.5 E_{F}$.}
\label{figS2}
\end{figure} 

However, the comparison reported in Fig.\ref{figS2} shows that, if one tries to match the theoretical DOS curve and the experimental data near the main structure at $0.5 E_{F}$ by rescaling the experimental data, one ends up with a tail of the DOS calculation (which decays rather slowly like $\omega^{-3/2}$) that considerably overestimates the wing of the experimental data. 
In addition, the DOS calculation misses the structure of the main peak at low energy.

A slow frequency tail resulted already in the theoretical approaches of Refs.[S1,S2] which did not include final-state effects.
It is only with the subsequent inclusion of final-state effects [S3,S4] (which make the tail of the rf spectra to decay faster like $\omega^{-5/2}$ and pile up the oscillator strength toward threshold at the same time) that one is able to reproduce \emph{not only} the main structure at $0.5 E_{F}$ \emph{but also} the tail of the experimental data, as shown explicitly in Fig.3 of the main text.

\vspace{1cm}
\noindent
{\bf References}

\vspace{0.2cm}
\noindent
\footnotesize{[S1] J. Kinnunen, M. Rodriguez, and P. T{\"o}rm{\" a}, Science {\bf 305}, 1131 (2004).}\\
\footnotesize{[S2] Y. He, Q. Chen, and K. Levin,  Phys. Rev. A {\bf 72}, 011602(R) (2005).}\\
\footnotesize{[S3] A. Perali, P. Pieri, and G. C. Strinati, Phys. Rev. Lett. {\bf 100}, 010402 (2008).}\\
\footnotesize{[S4]  P. Pieri,  A. Perali, and G. C. Strinati, Nature Phys. {\bf 5}, 736 (2009).} 


\begin{thebibliography}{99}
\bibitem{footnote-loss} We speculate that the free-atom peak is more sensitive to saturation effects. 
                                     This may explain the reduction of the peak height in Fig.1(b) and also lead to the 
                                     up-shift of the peak in Fig.3.
\end{thebibliography}
\end{document}